\documentstyle[prl,aps]{revtex}

\begin{document}
\draft

\twocolumn[
\widetext
\hfill SMC-PHYS-158

\hfill hep-ph/9807534

\centerline{\large\bf A dynamical origin of the mass hierarchy among}
\centerline{\large\bf  neutrinos, charged leptons, and quarks}
\vskip 10pt
\centerline{                Keiichi Akama and Kazuo Katsuura}
\centerline{\small\sl       Department of Physics, Saitama Medical College,
                Kawakado, Moroyama, Saitama, 350-0496, Japan}
\vskip 5pt

\leftskip 15mm\rightskip 15mm
\begin{abstract}
\baselineskip  = 10pt
We propose a dynamical mass-generation scenario
which naturally realizes the mass hierarchy among
the neutrinos, charged leptons and quarks,
where the mass is dominated by the self-mass induced
through the anomalous (i.e. non-minimal) gauge interactions.

\end{abstract}

]

\narrowtext

The zenith-angle dependence of the atmospheric muon-neutrino flux
recently reported by Super-Kamiokande Collaboration and by MACRO Collaboration
\cite{exp}
seems to provide a strong evidence for the neutrino oscillation \cite{nuos},
and hence for the neutrino mass,
which has long been suggested by
the phenomena of the solar neutrino deficit \cite{solar}, \cite{suggest}
and the atmospheric neutrino deficit \cite{atmo}, \cite{suggest2}.
Then an important theoretical question is
why the neutrinos have so tiny masses
in contrast with the other fundamental fermions
(charged leptons and quarks).
Another eminent feature of the mass spectrum is
that the quarks are heavier than
the charged leptons in the same generation.
The fermions seem to acquire their masses
according to their interaction activity.
The standard model
(the quantum chromodynamics and the quantum electroweak dynamics
of the three generations of the quarks and leptons)
never explains the origins of any characteristics of the mass spectrum,
since the masses are proportional to the Yukawa coupling constants
which are free parameters of the theory.
In general, fermion masses can be kept naturally (in 't Hooft's sense) small
due to protection by the chiral symmetry \cite{tHooft}
(the exception of the top quark is another problem).
The chiral symmetries for the neutrinos are very stringent,
while those for the others are comparatively loose.
What dynamical mechanism makes them so different?
The large mass discrepancy between the neutral and the charged fermions
suggests that the electromagnetic interactions are concerned with it.
Typical effects which break the chiral symmetry are
the masses and anomalous (non-minimal) gauge interactions
of the fermions.
Hence we expect that the anomalous interactions
play an important role in the mass generation of the fermions \cite{mass}.
In fact, the anomalous interaction gives rise to
a severely divergent self-mass to the charged fermion.
In this letter, we propose a dynamical mass-generation scenario
which naturally realizes the mass hierarchy
among the neutrinos, charged leptons and quarks,
where the mass is dominated by the self-mass induced
through the anomalous interactions.

The anomalous interactions may not be ultimately fundamental,
because theories with them are not renormalizable.
They should be taken as an induced effect
arising from the underlying fundamental theory.
Various candidates of such a underlying theory are extensively investigated in literature.
They include, for example, the composite model \cite{comp},
the (supersymmetric) grand unified model \cite{gut},
the technicolor model \cite{tc}, and the superstring model \cite{ss}.
Here we do not precisely specify what the underlying theory is.
We simply assume a slightly broken chiral symmetry
and existence of a fundamental mass scale $\Lambda $,
which serves as a momentum cutoff in the low energy effective theory.
As a result of the underlying dynamics,
small mass and anomalous interaction terms are expected to
arise in the action integral of the low energy effective theory.
They are taken as the bare masses and bare interactions
in the effective theory,
and their quantum effects below the scale $\Lambda $
should be taken into account.
In general, the bare terms originating from the underlying theory include
both dimensionless and dimensional ones,
where the dimension means the mass scale dimension
of the action integral apart from the coefficient.
We define the dimensionless coupling constant of dimensional interaction
by scaling it by the fundamental mass scale $\Lambda $, and
we assume that thus defined dimensionless coupling constants
are less than unity
in order for the perturbation expansion to be meaningful.
The higher order diagrams in the dimensional interaction
diverge severely, but, at the same time,
are suppressed by inverse powers of $\Lambda $ in the coupling constants,
so that, as a whole, these diagrams behave like $\Lambda ^d$,
where $d$ is the scale dimension of the diagram.
Among these higher order diagrams,
those with negative dimension ($d<0$)
are suppressed by inverse powers of $\Lambda $,
and negligibly small,
while those with non-negative dimension ($d\ge 0$) diverge with $\Lambda $.
The corrections to
masses, which have positive dimension,
are enhanced by powers of $\Lambda $,
and those to dimensionless interactions
are enhanced by powers of $\ln\Lambda $.
These behaviors are no worse than those in the renormalizable theory.
The divergent parts belong to the same primitively divergent parts
as in the renormalizable theory,
and are absorbed by the renormalized masses and coupling constants
which are independently determined by experiments.
Thus, in the limit $\Lambda \rightarrow \infty $
the non-renormalizable dimensional interactions have
no observable effects, and
the effective theory looks like renormalizable (in power counting)
at low energies.
The renormalization theory by itself has no predicative power on the masses
and coupling constants because they are to be renormalized.
In the effective theory, however,
the fundamental scale $\Lambda $ is finite, and
the bare masses and the bare dimensionless interactions
are determined by the underlying theory in principle,
so that the enhanced contributions to them from
the non-renormalizable interactions acquire physical meaning.
Even though the fundamental theory is not precisely known,
knowledge on symmetries, quantum numbers etc.\
can make them observable.
The mass spectrum can be a good place to seek for the new physics signatures
arising from the non-renormalizable interactions,
as well as their direct effects.

We first consider the case where the lepton or quark field $\psi $
with the electric charge $Q$
interacts with the photon field $A_\mu $.
We assume that the chiral symmetry for fermions is slightly broken
in the underlying theory.
Then it gives rise to a small bare mass term and
a small bare anomalous electromagnetic interaction term
in the Lagrangian of the effective theory:
\begin{eqnarray}
\overline \psi \left( -m_0+{1\over 2}\mu _0\sigma ^{\mu \nu }F_{\mu \nu }\right) \psi ,
\end{eqnarray}
where $m_0$ and $\mu _0$ are the induced bare mass and
the bare anomalous magnetic moment, respectively,
and $F_{\mu \nu }=\partial _\mu A_\nu -\partial _\nu A_\mu $ is
the electromagnetic field strength.
The assumption of the slightly broken chiral symmetry implies that
\begin{eqnarray}
m_0\ll \Lambda , \ \ \ \ |\mu _0|\ll 1/\Lambda .  \label{csb}
\end{eqnarray}
In other words, the mass $m_0$ and the anomalous magnetic moment $\mu _0$
can be naturally small due to the protection by the chiral symmetry.
The unknown underlying dynamics determines the relative importance
of $m_0/\Lambda $ and $\mu _0\Lambda $.
Some symmetry or dynamics may suppress or enhance one or another of them.
Consider the simultaneous operation of
the discrete chiral transformation of the fermion field $\psi $
and the sign change of the photon field $A_\mu $:
\begin{eqnarray}
\cases{\psi \rightarrow \psi '=\gamma _5\psi ,\cr
A_\mu \rightarrow A'_\mu =-A_\mu .}  \label{SYM}
\end{eqnarray}
The anomalous electromagnetic interaction is invariant under (\ref{SYM}),
while the mass term is not.
The minimal electromagnetic interaction also violates
the invariance under (\ref{SYM}).
Suppose we first switch off the minimal electromagnetic interaction
($e\rightarrow 0$ where $e$ is the minimal electromagnetic coupling constant)
and respect the symmetry under (\ref{SYM}),
so that the anomalous magnetic moment is allowed,
but the mass is not.
Then we switch on the minimal electromagnetic interaction,
so that the symmetry under (\ref{SYM}) is broken
and the mass is generated.
The mass thus generated is expected to be proportional to $eQ$,
where $Q$ is the electric charge of the fermion field.
If we assume a slightly broken symmetry under (\ref{SYM})
in the limit $e\rightarrow 0$, we have
\begin{eqnarray}
m_0/\Lambda \ll |\mu _0|\Lambda   \label{m0mu0}
\end{eqnarray}
In other words, the mass $m_0$ can be naturally small
due to the protection by the discrete symmetry
under transformation (\ref{SYM})
in the limit $e\rightarrow 0$.

\def\ggg{
\special{pa  500  -0.0}%
\special{pa  -500   -0.0}%
\special{fp}%
\special{pa    275      0}%
\special{pa    273     -4}%
\special{pa    270     -8}%
\special{pa    264    -12}%
\special{pa    257    -16}%
\special{pa    249    -19}%
\special{pa    241    -22}%
\special{pa    233    -25}%
\special{pa    227    -28}%
\special{pa    223    -31}%
\special{pa    222    -35}%
\special{pa    222    -38}%
\special{pa    225    -43}%
\special{pa    230    -47}%
\special{pa    236    -52}%
\special{pa    243    -58}%
\special{pa    249    -64}%
\special{pa    255    -69}%
\special{pa    259    -75}%
\special{pa    261    -80}%
\special{pa    261    -84}%
\special{pa    259    -88}%
\special{pa    254    -91}%
\special{pa    247    -93}%
\special{pa    239    -94}%
\special{pa    230    -95}%
\special{pa    222    -96}%
\special{pa    214    -96}%
\special{pa    207    -97}%
\special{pa    203    -99}%
\special{pa    200   -102}%
\special{pa    199   -105}%
\special{pa    201   -110}%
\special{pa    204   -116}%
\special{pa    208   -123}%
\special{pa    213   -130}%
\special{pa    217   -138}%
\special{pa    221   -145}%
\special{pa    223   -151}%
\special{pa    223   -157}%
\special{pa    222   -161}%
\special{pa    218   -164}%
\special{pa    213   -165}%
\special{pa    206   -165}%
\special{pa    198   -164}%
\special{pa    190   -162}%
\special{pa    181   -160}%
\special{pa    174   -158}%
\special{pa    167   -157}%
\special{pa    162   -157}%
\special{pa    159   -159}%
\special{pa    157   -162}%
\special{pa    157   -167}%
\special{pa    158   -174}%
\special{pa    160   -181}%
\special{pa    162   -190}%
\special{pa    164   -198}%
\special{pa    165   -206}%
\special{pa    165   -213}%
\special{pa    164   -218}%
\special{pa    161   -222}%
\special{pa    157   -223}%
\special{pa    151   -223}%
\special{pa    145   -221}%
\special{pa    138   -217}%
\special{pa    130   -213}%
\special{pa    123   -208}%
\special{pa    116   -204}%
\special{pa    110   -201}%
\special{pa    105   -199}%
\special{pa    102   -200}%
\special{pa     99   -203}%
\special{pa     97   -207}%
\special{pa     96   -214}%
\special{pa     96   -222}%
\special{pa     95   -230}%
\special{pa     94   -239}%
\special{pa     93   -247}%
\special{pa     91   -254}%
\special{pa     88   -259}%
\special{pa     84   -261}%
\special{pa     80   -261}%
\special{pa     75   -259}%
\special{pa     69   -255}%
\special{pa     64   -249}%
\special{pa     58   -243}%
\special{pa     52   -236}%
\special{pa     47   -230}%
\special{pa     43   -225}%
\special{pa     38   -222}%
\special{pa     35   -222}%
\special{pa     31   -223}%
\special{pa     28   -227}%
\special{pa     25   -233}%
\special{pa     22   -241}%
\special{pa     19   -249}%
\special{pa     16   -257}%
\special{pa     12   -264}%
\special{pa      8   -270}%
\special{pa      4   -273}%
\special{pa      0   -274}%
\special{pa     -4   -273}%
\special{pa     -8   -270}%
\special{pa    -12   -264}%
\special{pa    -16   -257}%
\special{pa    -19   -249}%
\special{pa    -22   -241}%
\special{pa    -25   -233}%
\special{pa    -28   -227}%
\special{pa    -31   -223}%
\special{pa    -35   -222}%
\special{pa    -38   -222}%
\special{pa    -43   -225}%
\special{pa    -47   -230}%
\special{pa    -52   -236}%
\special{pa    -58   -243}%
\special{pa    -64   -249}%
\special{pa    -69   -255}%
\special{pa    -75   -259}%
\special{pa    -80   -261}%
\special{pa    -84   -261}%
\special{pa    -88   -259}%
\special{pa    -91   -254}%
\special{pa    -93   -247}%
\special{pa    -94   -239}%
\special{pa    -95   -230}%
\special{pa    -96   -222}%
\special{pa    -96   -214}%
\special{pa    -97   -207}%
\special{pa    -99   -203}%
\special{pa   -102   -200}%
\special{pa   -105   -199}%
\special{pa   -110   -201}%
\special{pa   -116   -204}%
\special{pa   -123   -208}%
\special{pa   -130   -213}%
\special{pa   -138   -217}%
\special{pa   -145   -221}%
\special{pa   -151   -223}%
\special{pa   -157   -223}%
\special{pa   -161   -222}%
\special{pa   -164   -218}%
\special{pa   -165   -213}%
\special{pa   -165   -206}%
\special{pa   -164   -198}%
\special{pa   -162   -190}%
\special{pa   -160   -181}%
\special{pa   -158   -174}%
\special{pa   -157   -167}%
\special{pa   -157   -162}%
\special{pa   -159   -159}%
\special{pa   -162   -157}%
\special{pa   -167   -157}%
\special{pa   -174   -158}%
\special{pa   -181   -160}%
\special{pa   -190   -162}%
\special{pa   -198   -164}%
\special{pa   -206   -165}%
\special{pa   -213   -165}%
\special{pa   -218   -164}%
\special{pa   -222   -161}%
\special{pa   -223   -157}%
\special{pa   -223   -151}%
\special{pa   -221   -145}%
\special{pa   -217   -138}%
\special{pa   -213   -130}%
\special{pa   -208   -123}%
\special{pa   -204   -116}%
\special{pa   -201   -110}%
\special{pa   -199   -105}%
\special{pa   -200   -102}%
\special{pa   -203    -99}%
\special{pa   -207    -97}%
\special{pa   -214    -96}%
\special{pa   -222    -96}%
\special{pa   -230    -95}%
\special{pa   -239    -94}%
\special{pa   -247    -93}%
\special{pa   -254    -91}%
\special{pa   -259    -88}%
\special{pa   -261    -84}%
\special{pa   -261    -80}%
\special{pa   -259    -75}%
\special{pa   -255    -69}%
\special{pa   -249    -64}%
\special{pa   -243    -58}%
\special{pa   -236    -52}%
\special{pa   -230    -47}%
\special{pa   -225    -43}%
\special{pa   -222    -38}%
\special{pa   -222    -35}%
\special{pa   -223    -31}%
\special{pa   -227    -28}%
\special{pa   -233    -25}%
\special{pa   -241    -22}%
\special{pa   -249    -19}%
\special{pa   -257    -16}%
\special{pa   -264    -12}%
\special{pa   -270     -8}%
\special{pa   -273     -4}%
\special{pa   -274      0}%
\special{fp}%
}
\begin{figure}
\begin{center}
\begin{picture}(0,15)
\put(-60,-5){%
\put(-50,5){a}%
\special{pn 10}%
\ggg%
\special{pn 15}%
\special{pa 50  50}%
\special{pa -50 -50}%
\special{fp}%
\special{pa 50  -50}%
\special{pa -50 50}%
\special{fp}%
}
\put(60,-5){%
\put(-50,5){b}%
\special{pn 10}%
\ggg%
\special{pn 15}%
\special{pa 325  75}%
\special{pa 175 -75}%
\special{fp}%
\special{pa 325  -75}%
\special{pa 175 75}%
\special{fp}%
\special{sh 1}%
\special{ar 250 0 50 50 0 6.2832}%
}
\end{picture}
\end{center}
\caption{}
\label{fig1}
\end{figure}

Now we consider the quantum effects arising from the
bare mass and anomalous electromagnetic interaction
which originally stem from the underlying theory.
At the one-loop level, the contributions to the mass
come from the diagrams in Fig.\ \ref{fig1}a and b.
The solid and the wavy lines indicate
the fermion and the photon propagators, respectively,
and the simple and the blobbed vertices indicate the minimal
and the anomalous electromagnetic interactions, respectively.
The contribution to the fermion mass takes place only
when the chiral symmetry is broken in the course of the diagram.
The cross indicates the part which breaks the chiral symmetry.
In diagram a, the fermion propagator breaks the chiral symmetry through the
bare mass $m_0$,
and hence the amplitude is proportional to $m_0$.
In diagram b, the anomalous electromagnetic interaction vertex
breaks the chiral symmetry,
and the amplitude is proportional
to the anomalous magnetic moment $\mu _0$.
Diagram a diverges logarithmically and diagram b quadratically.
We incorporate the momentum cutoff at the fundamental scale $\Lambda $
by inserting the form factor
\begin{eqnarray}
1/(1-q^2/\Lambda ^2)^2
\end{eqnarray}
to the photon line with the momentum $q$,
and nothing to the fermion line at the one-loop level,
so that it makes the integral sufficiently convergent and,
at the same time,
the cutoff preserves the gauge symmetry and the chiral symmetry.
The cutoff should preserve also the chiral symmetry, which is broken,
in order to guarantee
that the breaking takes place only through dynamics.
Then it is straightforward to obtain the result for the mass:
\begin{eqnarray}
m=m_0+{3\over 16\pi ^2}e^2Q^2m_0\ln{\Lambda ^2\over m_0^2}
-{3\over 8\pi ^2}eQ\mu _0\Lambda ^2
\label{m}
\end{eqnarray}
where we have neglected the term suppressed by inverse powers of $\Lambda $.
The second term, which comes from diagram a,
is the usual self-mass in the quantum electrodynamics,
and is smaller than the bare term $m_0$
even if the cut off $\Lambda $ is as large as the Planck mass.
On the other hand, the third term, which is caused by
the anomalous electromagnetic interaction (diagram b),
is enhanced by the severely divergent nature of the diagram.
The sign of the third term depends on the signs of $Q$ and $\mu _0$.
The negative sign in the fermion mass can always be absorbed
by redefinition of the fermion field.
What is observable is their relative signs.
The absolute value of the third term
is much larger than the first term under the assumption (\ref{m0mu0}).
Because it is proportional to the electric charge,
it exists only for the charged leptons and quarks,
and is absent for the neutrinos.
{This fact explains the origin of the large mass discrepancy
between the neutrinos and the other charged fermions}.
Then the neutrino mass $m_\nu $ and the charged fermion mass $m_f$
are essentially given by
\begin{eqnarray}
&  & m_\nu =m_0,      \label{mnu}
\\&  & m_f={3\over 8\pi ^2}e|Q\mu _0|\Lambda ^2.  \label{mf}
\end{eqnarray}
The last relation (\ref{mf}) indicates that
the intrinsic anomalous magnetic moment is proportional to the mass,
contrary to the Schwinger correction which is
proportional to the mass inverse.
Another important difference between the intrinsic anomalous magnetic moment
arising from the fundamental dynamics
and that induced through ordinary quantum electrodynamics
(Schwinger correction) is as follows.
The former (intrinsic one) is a constant
as a function of the momenta, and persists up to the scale $\Lambda $,
while the latter diminishes with increasing momenta.
These properties are directly testable
by precision experiments of the
anomalous magnetic moment of leptons \cite{mueexp}--\cite{mumuth},
cross sections in lepton-(anti)lepton scattering,
scaling violation in deep inelastic
lepton-hadron scattering \cite{deep}, etc.\ in the future.
At present the experimental result for
the anomalous magnetic moment \cite{mueexp}
of the electron $\mu _e$ agree well
with the theoretical prediction from
the quantum electrodynamics \cite{mueth}
within errors, which sets the stringent limit
\begin{eqnarray}
-2.0\times 10^{-5}{\rm TeV}^{-1}<\mu _e<1.3\times 10^{-5}{\rm TeV}^{-1}.
\end{eqnarray}
If we apply this to the relation (\ref{mf}),
we obtain the lower bound on the compositeness scale $\Lambda $ as
\begin{eqnarray}
\Lambda >2{\rm TeV}.
\end{eqnarray}
Similarly, the agreement between the experimental result \cite{mumuexp} and
the quantum-electrodynamics prediction \cite{mumuth}
on the anomalous magnetic moment $\mu _\mu $ of the muon
sets the limit
\begin{eqnarray}
-2.0\times 10^{-5}{\rm TeV}^{-1}<\mu _e<3.1\times 10^{-5}{\rm TeV}^{-1},
\end{eqnarray}
which, if applied to (\ref{mf}), gives the lower bound
\begin{eqnarray}
\Lambda >15{\rm TeV}.
\end{eqnarray}
It is also noteworthy that the top quark may have comparatively large
intrinsic anomalous magnetic moment.
If we dare to speculate,
it is plausible that $m_0$ in (\ref{mnu})
is common to the neutrino species,
so that they are almost degenerate,
giving chances to large mixing
which is suggested by the observations
on the solar and atmospheric neutrinos.

Let us incorporate the weak interactions into this scheme.
In general the weak bosons can also have anomalous interactions:
\begin{eqnarray}& & { }{ }{ }
\frac{1}{2}{\sum_{i=1}^2}\left[ 
\overline \psi _{i}\sigma ^{\mu \nu }
(\mu _iF_{\mu \nu }+\mu '_iZ_{\mu \nu })\psi _{i}\right] 
\cr &  &+{1\over \sqrt 2}
\overline \psi _{1}\sigma ^{\mu \nu }W_{\mu \nu }^+
(\mu ''_1\gamma _L-\mu ''_2\gamma _R)\psi _{2}
+{\rm h.c.},
\label{AWZ}
\end{eqnarray} 
where 
$\psi _1$ and $\psi _2$ stand for up-type and down-type fermions, 
respectively,
$Z_{\mu \nu }$ and $W_{\mu \nu }$ are the field strength of the 
$Z$ and $W$ bosons, respectively,
$\gamma _R=(1+\gamma _5)/2$, $\gamma _L=(1-\gamma _5)/2$,
and $\mu _i$, $\mu '_i$, and $\mu ''_i(i=1,2)$ 
are anomalous coupling constants.
In the context of the electroweak $SU(2)_L\times U(1)_Y$ gauge theory,  
the anomalous interactions are introduced 
th
rough the non-renormalizable interactions
\begin{eqnarray}
\overline \psi _{iR}\phi ^{(c)\dagger }i\sigma ^{\mu \nu }
(\kappa _iB_{\mu \nu }+\kappa '_iW^a_{\mu \nu }\tau _a)\psi _L+{\rm h.c.},
\label{AEW}
\end{eqnarray}
where
$\kappa _i$ and $\kappa '_i(i=1,2)$ are the coupling constants,
$\psi _{iR}(i=1,2)$ is the $SU(2)_L$-singlet righthanded component of
the quark (or lepton) field $\psi _i$,
$\psi _L=(\psi _{1L},\psi _{2_L})$
is the $SU(2)_L$
-doublet lefthanded component of $\psi =(\psi _1,\psi _2)$,
$\phi ^{(c)}$ is the $SU(2)_L$-doublet Higgs field for $i=2$
or its charge conjugate for $i=1$,
$B_{\mu \nu }$ and $W^a_{\mu \nu }(a=1,2,3)$ are the field strengths
of the electroweak gauge fields,
and $\tau _a(a=1,2,3)$ is the Pauli matrix.
As is argued above, such non-renormalizable interactions
can exist in the effective theory
as far as they are suppressed by the inverse of
the fundamental scale $\Lambda $.
When the  $SU(2)_L\times U(1)_Y$ gauge symmetry is spontaneously broken,
the neutral component of the Higgs field
acquires a non-vanishing vacuum expectation value,
and (\ref{AEW}) gives rise to the anomalous interaction (\ref{AWZ})
with the constraint
\begin{eqnarray}
\mu ''_i=\mu _i\sin\theta _W+\mu '_i\cos\theta _W,\ \ \ (i=1,2)
\end{eqnarray}
where $\theta _W$ is the Weinberg angle.
Then the self-mass to be added to (\ref{m}) is calculated to be
\begin{eqnarray}
-{3\over 16\pi ^2}\left[ \mu 'g\left( \pm {1\over 2}-2Q\sin^2\theta _W\right)
\pm \mu ''g\right] \Lambda ^2,\label{mfEW}
\end{eqnarray}
where $g$ is the gauge coupling constant of $SU(2)_L$,
and $\pm $ is $+(-)$ for the up(down)-type fermions.
In the standard model the right-handed component of the neutrinos $\nu _R$
has no interaction.
It is natural to assume that the anomalous interactions of $\nu _R$
are also absent, which means that $\mu =\mu '=\mu ''=0$
for neutrinos.
Then the self-mass in (\ref{mfEW}) for the neutrinos is absent,
and the mass discrepancy with the others remains large.

It is straightforward to incorporate the strong interaction
to the present scheme
by combining it with the quantum chromodynamics.
The growth of the running gauge coupling constant at low energies
may affect the masses and the anomalous interactions of quarks,
and the proportionality between them
in (\ref{mf}) and (\ref{mfEW}) may be distorted.
Furthermore the slight breaking of the chiral symmetry
in the fundamental dynamics gives rise to
the small anomalous chromomagnetic interactions
\begin{eqnarray}
{1\over 2}\mu _c\overline q\sigma ^{\mu \nu }G^a_{\mu \nu }\lambda _aq
  \label{AQC},
\end{eqnarray}
where $\mu _c$ is the anomalous chromomagnetic moment,
$q$ is the quark field,
$G_{\mu \nu }$ is the gluon field strength,
and $\lambda _a$ is the Gell-mann matrix for $SU(3)_c$.
In the context of the standard model,
the anomalous interaction (\ref{AQC}) can arise from
the non-renormalizable interactions
\begin{eqnarray}
\kappa _c\overline \psi _{Ri}\phi ^{(c)\dagger }i
\sigma ^{\mu \nu }G^a_{\mu \nu }\lambda _a\psi _L+{\rm h.c.}.
\end{eqnarray}
The diagram Fig.\ \ref{fig1}b with the interaction (\ref{AQC})
gives rise to the additional contributions
\begin{eqnarray}
-{1\over \pi ^2}g_s\mu _c\Lambda ^2 \label{mfQCD}
\end{eqnarray}
to the quark self-mass,
where $g_s$ is the gauge coupling constant of the quantum chromodynamics.
Since running coupling constant $g_s$ is much enhanced at low energies,
the term (\ref{mfQCD}) dominates over (\ref{mfEW}).
Qualitatively, it explains why the quarks have larger masses
than the charged lepton in the same generation.

In conclusion, the anomalous interactions
originating from the unknown fundamental dynamics can give rise to
a large self-mass of the fermions
according to their interaction activity,
which realizes the gross feature of
the mass hierarchy among neutrino, charged lepton and quarks
in a generation.
This scheme predicts that the mass
is proportional to the strength of the anomalous interactions.
In a toy model only with electromagnetism
the relation is simple and testable.
If we incorporate the quantum electroweak dynamics
and quantum chromodynamics,
the predictions become somewhat obscure
because of the additional free parameters,
though it is still testable in principle.
The flavor mixing occurs through the transition anomalous magnetic moment,
and it is related to the anomalous radiative decays of the fermions.
In the present scheme,
masses from the ordinary Higgs mechanism are
as tiny as the neutrino masses,
which means that the Yukawa coupling constants are as small
as those of the neutrinos.
Instead the radiative corrections of the Yukawa coupling constants
due to the anomalous interactions are enhanced
in proportion with their enhanced physical mass.
Thus we expect that similar pattern of the branching ratios of the Higgs
particle as that in the standard model.
Interesting phenomenological aspects including them
are now under investigation,  and will be presented elsewhere.

\end{document}